\documentclass{article}
\usepackage[compatibility=false]{caption}
\usepackage{lipsum} 
\usepackage{graphicx} 
\usepackage{float}
\usepackage{booktabs}
\usepackage{caption}
\usepackage{amssymb}
\usepackage{cite}
\usepackage{url}
\usepackage{adjustbox}
\usepackage{bm}
\usepackage{spconf,amsmath,amssymb,graphicx,hyperref}
\usepackage{multirow}
\usepackage{booktabs}
\usepackage{bm}
\usepackage[table]{xcolor}
\usepackage{enumitem}

\title{FUSEMOS: Perceptual Evaluation of Text-to-Music Generation with Dual-Encoder Fusion and Ranking-Aware Composite Loss}
\name{Jing Yang\textsuperscript{1,3}$^{\dagger}$, Haoyu Wang\textsuperscript{2,3}$^{\dagger}$, Ningning Pan\textsuperscript{2*}, Zhao Wang\textsuperscript{3}, Jianxuan Yang\textsuperscript{3}, Gongping Huang\textsuperscript{1*}
\thanks{$^{\dagger}$ Equal Contribution. * Corresponding Author.}
\thanks{This work was supported by the National Natural Science Foundation (NSFC) of China under Grant 62471340 and 62401479, and Sichuan Science and Technology Program under Grant 2026NSFSC1430.}
}
\address{
\textsuperscript{1}School of Electronic Information, Wuhan University, Wuhan Hubei, 430072,  China \\
\textsuperscript{2} School of Computing
and Artificial Intelligence, \\Southwestern University of Finance and Economics,
Chengdu, Sichuan 610074, China\\
\textsuperscript{3}MiLM Plus, Xiaomi Inc., Wuhan, China
}
\begin{document}
\ninept
\maketitle
\begin{abstract}
Perceptual evaluation of generated music is essential for developing text-to-music (TTM) generation systems. 
Existing automatic music perceptual evaluation methods rely solely on the contrastive language audio pretraining (CLAP) model to extract both audio and text features for predicting human-annotated mean opinion scores (MOS), particularly for overall musical impression (OMI) and textual alignment (TA). 
However, CLAP is limited in capturing fine-grained musical attributes such as timbre and expressiveness \emph{etc}. 
To address this, we propose FUSEMOS, a dual-encoder fusion framework that integrates CLAP with MERT, a self-supervised model pretrained on large scale music audio, to better capture both high level semantic alignment and detailed musical characteristics for more accurate OMI and TA estimation. 
Furthermore, to overcome the limitations of conventional regression losses in measuring human preferred relative ranking of music segments, we design a novel ranking-aware composite loss function.
Experiments conducted on the MusicEval benchmark demonstrate that our approach outperforms the existing SOTA methods in both mean squared error and rank correlation metrics by a large margin.

\end{abstract}
\section{Introduction}
\label{sec:intro}

Text-to-music (TTM) generation~\cite{zhang2025inspiremusic,prajwal2024musicflow,liu2024music,yan2023harmonizing} has advanced significantly in recent years.
A high-quality TTM system must simultaneously fulfill two essential requirements: (1) semantic alignment with the input text description, and (2) high musical quality across dimensions such as structure, expressiveness, and coherence.
Subjective evaluations, such as mean opinion scores (MOS), are considered the gold standard for assessing TTM generation systems. 
However, subjective tests are labor-intensive, costly, and difficult to reproduce, which hinders their scalability for iterative development and large-scale benchmarking.
On the other hand, objective metrics like Fréchet audio distance (FAD)~\cite{kilgour2019frechet} and inception score (IS) are computationally efficient and reproducible, but often exhibit weak correlation with human preferences \cite{vinay2022evaluating}.
Therefore, developing reliable and scalable perceptual evaluation methods is crucial for continued progress in TTM generation.

Automated prediction~\cite{mittag2021nisqa,wang2024uncertainty,qi2023ssl,shi2024samos} of human-annotated MOS for perceptual evaluation has recently gained research attention. Most existing methods, however, focus solely on assessing audios quality and neglect textual alignment (TA)~\cite{yao2025songeval, wisnu2025improving, tjandra2025meta}, which is inadequate for TTM generation systems. To address this gap, MusicEval~\cite{liu2025musiceval} introduced a benchmark dataset with human-rated MOS for both overall musical impression (OMI) and TA, enabling systematic and comprehensive evaluation of TTM models. 
OMI quantifies the overall quality of generated music, including aspects such as melody, rhythm, and expressiveness. TA measures the degree to which the musical content aligns with the input text.
The current benchmark leverages the contrastive language-audio pretraining (CLAP) model~\cite{wu2023large} to extract joint textual and audio features: an MLP head is applied to audio embeddings for OMI prediction, while concatenated audio-text features are used for TA regression.
Nevertheless, music perception encompasses a broad range of attributes, including timbre, rhythm, harmony, and structural coherence, that may not be adequately captured by a general-purpose audio encoder like CLAP, which is primarily optimized for cross-modal alignment. 
As a result, despite its strengths in semantic matching, CLAP may fail to represent intrinsic musical structures effectively when used in isolation.

To address these limitations, we propose FUSEMOS, a dual-encoder fusion framework with ranking-aware loss for TTM perceptual evaluation. Our approach leverages two complementary pretrained models: CLAP, which provides strong audio-text semantic alignment, and MERT~\cite{li2023mert}, a self-supervised model pretrained on large-scale music data that captures rich musical structure. Instead of fusing audio features early (e.g., via concatenation), we adopt a late fusion strategy for both OMI and TA scores prediction through dedicated mapping networks
allowing the model to exploit the distinct strengths of each representation path.
Besides, traditional regression objectives such as mean squared error (MSE) are sensitive to absolute score deviations, yet fail to capture the ordinal nature of human perceptual judgments, that is, the inherent ranking and relative preference among different music segments. 
Thus, we introduce an ranking-aware composite loss function that combines a truncated regression loss and a contrastive ranking loss. Experiments on the MusicEval benchmark show that FUSEMOS achieves higher correlation with human judgments than the single-encoder baselines. Ablation studies further validate the effectiveness of each component: the dual-encoder architecture captures complementary semantic and musical features, the late fusion strategy outperforms early fusion alternatives, and the proposed composite loss improves both regression accuracy and rank consistency.

\section{PROPOSED METHOD}
\label{sec:PROPOSED METHOD}

\begin{figure*}[t!]
    \centering
    \includegraphics[width=\linewidth]{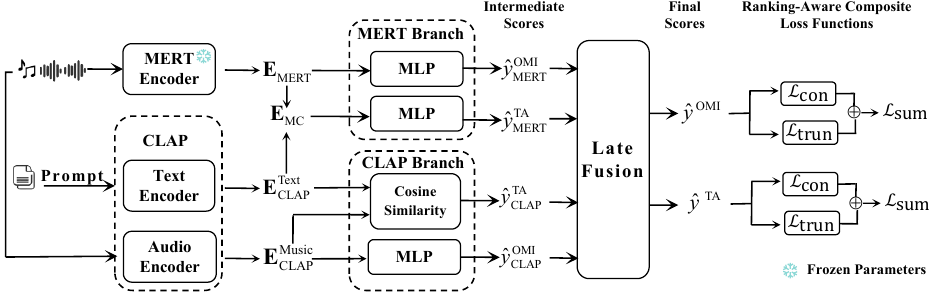 }
    \caption{The architecture of the proposed FUSEMOS framework for TTM perceptual evaluation. The model is trained with a ranking-aware composite loss that combines truncated regression ($\mathcal{L}_\text{trun}$) and contrastive ranking ($\mathcal{L}_\text{con}$) objectives.}
    \label{fig:model_arch}
\end{figure*}
Given a generated music signal and its corresponding text description, our goal is to predict two perceptual quality scores: OMI score, $\hat{y}^{\text{OMI}}$, and the TA score, $\hat{y}^{\text{TA}}$. As illustrated in Fig.~\ref{fig:model_arch}, our proposed framework FUSEMOS achieves this through a dual-encoder fusion architecture that leverages complementary representations from CLAP and MERT, with an innovative ranking-aware composite loss.

\subsection{Dual-Encoder Fusion Architecture}
\label{ssec:fusion}

\subsubsection{CLAP Branch}
The CLAP model is a dual-encoder framework trained on large-scale audio-text pairs to align semantically similar contents of audio and text in a shared embedding space. Given a generated music waveform and its corresponding text prompt, CLAP produces two $L_2$-normalized embeddings: $\mathbf{E}_{\text{CLAP}}^{\text{Text}}$, $\mathbf{E}_{\text{CLAP}}^{\text{Music}} \in \mathbb{R}^{ D_1}$, where $D_1$ is the embedding dimension. Both vectors reside in a common semantic space, enabling direct comparison via dot product.
Thanks to contrastive learning, CLAP excels at capturing semantic alignment between music signal and text, making it naturally suitable for predicting TA. 
Thus a coarse estimation of TA score can be computed as the cosine similarity between the text and music embeddings:
\begin{equation}
    \hat{y}_{\text{CLAP}}^{\text{TA}}  = \frac{
        \mathbf{E}_{\text{CLAP}}^{\text{Text}} \cdot 
        \mathbf{E}_{\text{CLAP}}^{\text{Music}}
    }{
        \left\| \mathbf{E}_{\text{CLAP}}^{\text{Text}} \right\|
        \left\| \mathbf{E}_{\text{CLAP}}^{\text{Music}} \right\|
    }.
\end{equation}
For coarse OMI estimation, which reflects general music quality rather than semantic content, we use the music embedding  $\mathbf{E}_{\text{CLAP}}^{\text{Music}}$ as input to a dedicated MLP:
\begin{equation}
    \hat{y}_{\text{CLAP}}^{\text{OMI}} = \text{MLP}^{\text{OMI}}_{\text{CLAP}}\left( \mathbf{E}_{\text{CLAP}}^{\text{Music}} \right).
\end{equation}

\subsubsection{MERT Branch}
MERT is a self-supervised audio representation model pretrained on large-scale music data, designed to capture fine-grained musical structure such as timbre, rhythm, and harmonic progression. 
We extract the hidden states from the final layer of MERT, denoted as $\mathbf{E}_0 \in \mathbb{R}^{ 1 \times T \times D_2}$, where $T$ is the number of time steps, and $D_2$ is the feature dimension. A learnable query-based attention pooling mechanism is then applied to obtain the final music representation $\mathbf{E}_{\text{MERT}} \in \mathbb{R}^{D_2}$:
\begin{equation}
    \mathbf{E}_{\text{MERT}} = \text{MultiheadAttn}(\mathbf{q}, \mathbf{E}_0, \mathbf{E}_0),
\end{equation}
where $\mathbf{q} \in \mathbb{R}^{1 \times 1 \times D_2}$ is a learnable query vector. This attention-based pooling enables the model to dynamically focus on perceptually important musical segments during representation aggregation.

Since the CLAP text encoder and the MERT model are trained independently and do not share a common representation space for text, we refrain from computing cosine similarity between them. Instead, we concatenate the music embedding $\mathbf{E}_{\text{MERT}}$ with the CLAP text embedding $\mathbf{E}_{\text{CLAP}}^{\text{Text}}$ to form a joint feature:
\begin{equation}
    \mathbf{E}_{\text{MC}} = (\mathbf{E}_{\text{MERT}} \oplus \mathbf{E}_{\text{CLAP}}^{\text{Text}}) \in \mathbb{R}^{ (D_1 + D_2)},
\end{equation}
where $\oplus$ denotes feature concatenation. The resulting features are passed through dedicated MLP to produce another intermediate estimations of OMI and TA scores:
\begin{align}
    \hat{y}_{\text{MERT}}^{\text{OMI}} &= \mathrm{MLP}^{\text{OMI}}_{\text{MERT}}(\mathbf{E}_{\text{MERT}}), \\
    \hat{y}_{\text{MERT}}^{\text{TA}} &= \mathrm{MLP}^{\text{TA}}_{\text{MERT}}\left[ \mathrm{LayerNorm}(\mathbf{E}_{\text{MC}}) \right].
\end{align}

\subsubsection{Late Fusion with Modality-Specific Mapping}
\label{ssec:mapping}
To fully exploit the complementary strengths of CLAP and MERT, we adopt a late fusion strategy that combines predictions from both encoders at the score level, rather than fusing raw features early in the network. This approach offers greater flexibility by allowing each encoder pathway to independently model perceptual aspects aligned with its inductive bias, before integrating their outputs into a unified prediction. The intermediate predictions, representing each branch’s preliminary assessment of perceptual quality, are concatenated and passed through a lightweight linear fusion head to produce the final scores:
\begin{align}
    \hat{y}^{\text{OMI}} &= \text{Linear}_{\text{OMI}}\left[(\hat{y}_{\text{CLAP}}^{\text{OMI}} \oplus \hat{y}_{\text{MERT}}^{\text{OMI}})\right], \\
     \hat{y}^{\text{TA}} &= \text{Linear}_{\text{TA}}\left[(\hat{y}_{\text{CLAP}}^{\text{TA}} \oplus \hat{y}_{\text{MERT}}^{\text{TA}}]\right).
\end{align}
The fusion head adaptively weights contributions from each branch. Unlike early fusion, which risks interference between heterogeneous representations, our approach preserves modality-specific integrity while enabling effective score-level integration.

\subsection{Ranking-Aware Composite Loss Functions}
\label{loss functions}
Effective training of FUSEMOS requires an objective function that not only minimizes prediction error but also captures the relative preferences in human judgments. Standard regression losses such as MSE focus on absolute errors but may fail to preserve the relative ranking of MOS ratings. To address this, we introduce a composite loss that combine contrastive ranking loss~\cite{saeki2022utmos},~\cite{serra2021sesqa} and truncated regression loss~\cite{leng2021mbnet} to improve ranking consistency with human perceptual behavior.

\subsubsection{ Contrastive Ranking Loss}
The contrastive ranking loss~\cite{saeki2022utmos},~\cite{serra2021sesqa}, denoted as $\mathcal{L}_\text{con}$, explicitly models the relative quality ranking relationships between samples by constraining the consistency between the predicted score differences and the true score differences of sample pairs, enhancing the model's discriminative. Initially, we define the difference measure $d_{ij}$ between sample $i$ and $j$ as follows:
\begin{align}
d_{ij} =& \left| (y_i - y_j) - (\hat{y}_i - \hat{y}_j) \right| - \alpha,
\end{align}
where $y_i$ and $y_j$ represent the true scores of samples $i$ and $j$, respectively, $\hat{y}_i$ and $\hat{y}_j$ represent the predicted scores by the model, and $\alpha$ is a hyperparameter that introduces a margin of tolerance in the difference measure. Then, the contrastive ranking loss function is calculated as:
\begin{align}
\mathcal{L}_\text{con} =& \frac{1}{N(N-1)} \sum_{i=1}^{N} \sum_{\substack{j=1 \\ j \neq i}}^{N} \max\left(0, d_{ij} \right), 
\end{align}
where $N$ represents the number of samples in a batch. Note that, $\max\left(0, d_{ij} \right)$ ensures that when the differences between the ground-truth score and the estimated score exceed $\alpha$ will be considered in the loss calculation, which are designed to avoid over-penalizing the model for small errors, thereby enhancing its robustness and generalization capability.

\subsubsection{Truncated Regression Loss}
The truncated regression loss~\cite{leng2021mbnet}, denoted as $\mathcal{L}_\text{trun}$, can prevent the model from overfitting to noisy labels or exact rating values by ignoring errors below a certain threshold $\tau$, which is defined as:
\begin{align}
\mathcal{L}_\text{trun} = \frac{1}{N} \sum_{i=1}^{N} \mathbb{I}\left(|y_i - \hat{y}_i| > \tau\right) \cdot |y_i - \hat{y}_i|,
\end{align}
where $\mathbb{I}(\cdot)$ denotes the indicator function, which equals $1$ if the condition is satisfied and $0$ otherwise.
By integrating contrastive ranking loss with truncated regression loss, the model can learn the relative quality of samples while maintaining the stability of the training process and preventing overfitting, thereby enhancing the model's generalization capability. 

The final total loss, $\mathcal{L}_\text{sum}$, is defined as:
\begin{align}
\mathcal{L}_\text{sum} = \beta \cdot \mathcal{L}_\text{con} + \gamma \cdot \mathcal{L}_\text{trun},
\end{align}
where $\beta$ and $\gamma$ are hyper parameters denoting weighting coefficients of the contrastive ranking loss and the truncated regression loss, respectively.

\section{EXPERIMENTS}
\label{sec:EXPERIMENTS}
\subsection{Dataset and Experimental Setup}
\label{sec:Implementation Details}

Experiments are conducted on MusicEval~\footnote{ \url{https://www.aishelltech.com/AISHELL_7A}}, the first expert-annotated dataset for the automatic evaluation of TTM systems in terms of both OMI and TA. 
There are $2748$ music clips generated by $31$ different models~\cite{majumder2024tango,zhu2023ernie,liu2024audioldm,ziv2024masked} under $384$ diverse text prompts. 
\begin{table*}[!ht]
    \centering
    \caption{Performance comparison of FUSEMOS, the baseline model, and its ablated variants on the MusicEval.}
    \begin{adjustbox}{width=\textwidth}
    \begin{tabular}{ccccccccc}
    \toprule
    & \multicolumn{8}{c}{\textbf{Overall musical impression}}\\
    \midrule
Model & U-MSE$\downarrow$ & U-LCC$\uparrow$& U-SRCC$\uparrow$ & U-KTAU$\uparrow$& S-MSE$\downarrow$& S-LCC$\uparrow$& S-SRCC$\uparrow$&S-KTAU$\uparrow$\\ \midrule
MusicEval (CLAP-only baseline) & $0.560$ & $0.637$& $0.637$ & $0.470$&$0.217$&$ 0.896$& $0.862$& $0.687$\\
FUSEMOS (Ours) & $\bm{0.260}$ & $\bm{0.810} $& $\bm{0.811}$ & $\bm{0.636}$ & $\bm{0.027}$ & $\bm{0.983}$ & $\bm{0.977}$ & $\bm{0.894}$ \\ \midrule
MERT-only & $0.45$ & $0.671$ & $0.669$ & $0.48$9 & $0.164$ & $0.936$ & $0.939$ & $0.807$ \\
FUSEMOS-$L_1$ & $0.268$ &$ 0.803$ & $0.797$  & $0.62$ & $0.029$ & $0.975$ & $0.965$ & $0.867$  \\
FUSEMOS-concat & $0.301 $& $0.786$ & $0.785$ &$ 0.602$ & $0.047$ & $0.973$ & $0.965$ & $0.862$ \\
FUSEMOS-atten & $0.402$ & $0.725$ & $0.726$ &$ 0.544$ & $0.136$ & $0.959$ & $0.949$ & $0.816$ \\
w/o-cos & $0.29$ & $0.784$ &$ 0.786$ &$ 0.608$ & $0.044$ & $0.967$ & $0.964$ & $0.867$ \\
\bottomrule
& \multicolumn{8}{c}{\textbf{Textual    alignment }} \\ \midrule
Model & U-MSE$\downarrow$ & U-LCC$\uparrow$& U-SRCC$\uparrow$ & U-KTAU$\uparrow$& S-MSE$\downarrow$& S-LCC$\uparrow$& S-SRCC$\uparrow$&S-KTAU$\uparrow$\\ \midrule 
MusicEval (CLAP-only baseline) & $0.568$& $0.502$& $0.495$&$ 0.354$& $0.116$&$ 0.875$& $0.861$&$ 0.692$\\ 
FUSEMOS (Ours) & $\bm{0.392}$ & $\bm{0.64}$ & $\bm{0.644}$ & $\bm{0.472}$ & $\bm{0.032}$ & $\bm{0.944}$ & $\bm{0.940}$ & $\bm{0.826}$ \\ \midrule
MERT-only & $0.573$ & $0.507$ & $0.522$ & $0.368 $& $0.129$ & $0.834$ & $0.848$ & $0.646$\\
FUSEMOS-$L_1$ & $0.464$ & $0.570 $&$ 0.576 $&$ 0.418$ & $0.05$ &$ 0.904 $ & $0.92$ & $0.785$ \\
FUSEMOS-concat & $0.444$ & $0.596$ & $0.605$ & $0.438$ & $0.057$ & $0.919$ &$ 0.904$ & $0.762$ \\
FUSEMOS-atten & $0.479$ & $0.539$ & $0.551$ & $0.397$ & $0.075$ & $0.866$ & $0.859$ & $0.669 $\\
w/o-cos & $0.468$ &$ 0.570$ &$ 0.583$ & $0.426$ &$ 0.040$  &$ 0.926$ & $0.925$ & $0.771$ \\
\bottomrule
    \end{tabular}
     \end{adjustbox}
    \label{tab:1}
\end{table*}

We fine-tuned a music-specialized CLAP model, initialized from a pre-trained checkpoint~\footnote{ \url {https://huggingface.co/lukewys/laion_clap/blob/main/music_audioset_epoch_15_esc_90.14.pt}} to enhance cross-modal semantic alignment for TTM evaluation. 
The MERT model~\footnote{ \url{https://huggingface.co/m-a-p/MERT-v1-95M}} was frozen to preserve its pre-trained musical structure representations. 
All experiments were conducted using the official train/validation/test split to ensure direct comparability with the benchmark. To ensure experimental fairness and reproducibility, we re-implemented the baseline method and evaluated it on our local system.
The proposed FUSEMOS is trained on a single NVIDIA H800 GPU with a batch size of $4$ for training and $2$ for validation. 
We set the hyperparameters as follows: $\tau = 0.3$, $\alpha = 0.2$, $\beta = 0.5$ and $\gamma = 1.0$. The margins $\tau$ and $\alpha$ are chosen empirically, while $\beta$ and $\gamma$ are optimized via validation.The model is trained using stochastic gradient descent (SGD) with a learning rate of $5 \times 10^{-4}$ and momentum $0.9$. 
All MLP components consisted of three fully connected layers with LeakyReLU activation functions.

\subsection{Results}
\label{sec:results}

We adopt both utterance-level (U-) and system-level (S-) metrics for evalutaion, including MSE and rank correlation metrics such as linear correlation coefficient (LCC), Spearman's rank correlation coefficient (SRCC), and Kendall's tau (KTAU). 
Utterance-level metrics are evaluated for each individual audio-text
pair, while system-level metrics aggregate predictions across all
samples from the same generation model. 
As shown in Table~\ref{tab:1}, FUSEMOS outperforms the MusicEval baseline across all metrics. For OMI, U-MSE and S-MSE are substantially reduced by $53.6\%$
and $87.6\%$,
respectively, with U-SRCC reaching $0.811$ and S-SRCC of $0.977$. 
For TA, U-SRCC improves from $0.495$ to $0.644$ and S-SRCC from $0.861$ to $0.940$. 
These results demonstrate that our dual-encoder framework achieves superior alignment with human judgments in both musical quality and semantic fidelity.

\subsection{Ablation Studies}
\label{sec:Ablation Studies}

We conducted ablation studies focused on four key aspects to validate the effectiveness of our design choices: the dual-encoder architecture, the late fusion strategy, the ranking-aware composite loss and the cosine similarity mechanism for TA prediction in the CLAP branch. Results are reported in Table.~\ref{tab:1}.
\begin{itemize}
\item \textbf{Ablation of Dual-Encoder (MERT-only):} 
    To evaluate the necessity of dual-encoder, we use only MERT branch for prediction ($\hat{y}^{\text{OMI}} = \hat{y}_{\text{MERT}}^{\text{OMI}}$, $\hat{y}^{\text{TA}} = \hat{y}_{\text{MERT}}^{\text{TA}} $). This variant removes the semantic alignment strength of CLAP's audio encoder, which leads to a significant drop in both OMI and TA performance. 
    Note that, the baseline method, MusicEval, only uses CLAP model for prediction, which achieves the similar results as MERT-only model.
    This comparison results demonstrate the CLAP and MERT encoders provides complementary music features that are both helpful in music perceptual evaluation. All ablations are trained with $L_1$ loss to ensure fair comparison with FUSEMOS-$L_1$.

\item \textbf{Ablation of Late Fusion:} 
    We compare our modality-specific late fusion with two alternative fusion: (1) concatenation: the MERT and CLAP music embeddings are directly concatenated as ($(\mathbf{E}_{\text{MERT}} \oplus \mathbf{E}_{\text{CLAP}}^{\text{Music}})$ and $(\mathbf{E}_{\text{MERT}} \oplus \mathbf{E}_{\text{CLAP}}^{\text{Music}} \oplus \mathbf{E}_{\text{CLAP}}^{\text{Text}})$), then fed into MLP for OMI and TA prediction, denoted as FUSEMOS-concat; 
(2) attention: a learnable attention mechanism is applied to fuse $\mathbf{E}_{\text{CLAP}}^{\text{Music}}$ and $\mathbf{E}_{\text{MERT}}$ at the feature level, with the resulting representation used to predict both scores via the MERT-style head, denoted as FUSEMOS-atten. 
Both variants risk interfering with the distinct inductive biases of each encoder. 
Results show that preserving independent pathways until the score level enables more effective integration of complementary features.
\item \textbf{Ablation of Composite Loss (FUSEMOS-$L_1$):} 
    To validate the effectiveness of the proposed ranking-aware composite loss, we conduct an ablation study by replacing it with the standard $L_1$ loss. The resulting model shows increased MSE and reduced rank correlation, indicating degraded regression accuracy and preference ranking consistency. Furthermore, we observe that using only the truncated regression loss reduces the S-SRCC to 0.867 (OMI) and 0.888 (TA), while relying solely on the contrastive ranking loss maintains high S-SRCC (0.963 and 0.938, respectively) but leads to a significant increase in MSE (7.887 and 12.50). These results confirm that both components are indispensable: truncated regression ensures stability and accuracy, while contrastive ranking preserves ordinal preference, and their combination achieves optimal balance. 

\item \textbf{Ablation of the Cosine Similarity Mechanism (w/o-cos):}
    To evaluate the effectiveness of using cosine similarity for TA prediction in the CLAP branch, we replaced the direct cosine similarity computation with a learnable MLP mapping (similar to the MERT branch):
    \begin{align}
    \hat{y}_{\text{CLAP}}^{\text{TA}} &= \mathrm{MLP}^{\text{TA}}_{\text{CLAP}}\left[ \mathrm{LayerNorm}(\mathbf{E}_{\text{CLAP}}^{\text{Music}} \oplus \mathbf{E}_{\text{CLAP}}^{\text{Text}}) \right].
    \end{align}
    The results show that this modification leads to a significant degradation in performance across all metrics, 
     indicating that the original cosine similarity-based approach is more effective at capturing semantic alignment. This result validates that CLAP’s inherent design, which learns to align audio and text embeddings in a shared space, is better leveraged through direct similarity computation rather than learning a complex transformation.
     \end{itemize}

\section{CONCLUSION}
\label{sec:conclusion}
This paper proposes FUSEMOS, a dual-encoder fusion framework for perceptual evaluation of TTM generation with a novel ranking-aware loss. 
By integrating the semantic alignment strength of CLAP and the musical structure modeling capability of MERT, our method leverages a late fusion strategy with modality-specific mapping to independently predict OMI and TA scores. To better preserve the relative ranking of human perceptual preferences, we novelly introduce an ranking-aware composite loss that combines a contrastive ranking loss and a truncated regression loss, to enhance both prediction robustness and rank consistency. 
Experimental results on the MusicEval benchmark demonstrate that FUSEMOS outperforms single-encoder baselines in both regression accuracy and rank correlation metrics by a large margin, \emph{e.g.,} over $50\%$ and $80\%$ improvement in terms of U-MSE and S-MSE respectively. 
These results validate the effectiveness of leveraging complementary pretrained models and modeling relative preferences in human ratings.
Future work will explore more sophisticated fusion mechanisms and the integration of additional music encoders to further improve evaluation performance.

\bibliographystyle{ieeetr}
\bibliography{refs}

@string{atraslp  = "IEEE/ACM Trans. Audio, Speech, Lang. Process."}

@string{icassp   = "Proc. IEEE ICASSP"}

@article{zhang2025inspiremusic,
  title={InspireMusic: Integrating Super Resolution and Large Language Model for High-Fidelity Long-Form Music Generation},
  author={Zhang, Chong and Ma, Yukun and Chen, Qian and Wang, Wen and Zhao, Shengkui and Pan, Zexu and Wang, Hao and Ni, Chongjia and Nguyen, Trung Hieu and Zhou, Kun and others},
  journal={arXiv preprint arXiv:2503.00084},
  year={2025}
}

@inproceedings{yan2023harmonizing,
  title={Harmonizing AI-Generated Music: Integrating Symbolic and Audio Models for Text-to-Music Generation},
  author={Yan, Xiping and Ma, Chunmei and Hao, Yonghong},
  booktitle={2023 2nd International Conference on Artificial Intelligence and Intelligent Information Processing (AIIIP)},
  pages={129--132},
  year={2023},
  organization={IEEE}
}

@inproceedings{liu2024music,
  title={Music understanding llama: Advancing text-to-music generation with question answering and captioning},
  author={Liu, Shansong and Hussain, Atin Sakkeer and Sun, Chenshuo and Shan, Ying},
  booktitle=icassp,
  pages={286--290},
  year={2024},
  organization={IEEE}
}

@inproceedings{prajwal2024musicflow,
  title={MusicFlow: Cascaded Flow Matching for Text Guided Music Generation},
  author={Prajwal, KR and Shi, Bowen and Le, Matthew and Vyas, Apoorv and Tjandra, Andros and Luthra, Mahi and Guo, Baishan and Wang, Huiyu and Afouras, Triantafyllos and Kant, David and others},
  booktitle={International Conference on Machine Learning},
  pages={41052--41063},
  year={2024},
  organization={PMLR}
}

@inproceedings{liu2025musiceval,
  title={Musiceval: A generative music dataset with expert ratings for automatic text-to-music evaluation},
  author={Liu, Cheng and Wang, Hui and Zhao, Jinghua and Zhao, Shiwan and Bu, Hui and Xu, Xin and Zhou, Jiaming and Sun, Haoqin and Qin, Yong},
  booktitle=icassp,
  pages={1--5},
  year={2025},
  organization={IEEE}
}

@article{tjandra2025meta,
  title={Meta audiobox aesthetics: Unified automatic quality assessment for speech, music, and sound},
  author={Tjandra, Andros and Wu, Yi-Chiao and Guo, Baishan and Hoffman, John and Ellis, Brian and Vyas, Apoorv and Shi, Bowen and Chen, Sanyuan and Le, Matt and Zacharov, Nick and others},
  journal={arXiv preprint arXiv:2502.05139},
  year={2025}
}

@article{yao2025songeval,
  title={SongEval: A Benchmark Dataset for Song Aesthetics Evaluation},
  author={Yao, Jixun and Ma, Guobin and Xue, Huixin and Chen, Huakang and Hao, Chunbo and Jiang, Yuepeng and Liu, Haohe and Yuan, Ruibin and Xu, Jin and Xue, Wei and others},
  journal={arXiv preprint arXiv:2505.10793},
  year={2025}
}

@inproceedings{li2023mert,
  title={MERT: Acoustic music understanding model with large-scale self-supervised training},
  author={Yizhi, LI and Yuan, Ruibin and Zhang, Ge and Ma, Yinghao and Chen, Xingran and Yin, Hanzhi and Xiao, Chenghao and Lin, Chenghua and Ragni, Anton and Benetos, Emmanouil and others},
  booktitle={The Twelfth International Conference on Learning Representations},
  year={2023}
}

@inproceedings{qi2023ssl,
  title={LE-SSL-MOS: Self-supervised learning MOS prediction with listener enhancement},
  author={Qi, Zili and Hu, Xinhui and Zhou, Wangjin and Li, Sheng and Wu, Hao and Lu, Jian and Xu, Xinkang},
  booktitle={2023 IEEE Automatic Speech Recognition and Understanding Workshop (ASRU)},
  pages={1--6},
  year={2023},
  organization={IEEE}
}

@inproceedings{wu2023large,
  title={Large-scale contrastive language-audio pretraining with feature fusion and keyword-to-caption augmentation},
  author={Wu, Yusong and Chen, Ke and Zhang, Tianyu and Hui, Yuchen and Berg-Kirkpatrick, Taylor and Dubnov, Shlomo},
  booktitle=icassp,
  pages={1--5},
  year={2023},
  organization={IEEE}
}

@inproceedings{shi2024samos,
  title={SAMOS: A Neural MOS Prediction Model Leveraging Semantic Representations and Acoustic Features},
  author={Shi, Yu-Fei and Ai, Yang and Lu, Ye-Xin and Du, Hui-Peng and Ling, Zhen-Hua},
  booktitle={2024 IEEE 14th International Symposium on Chinese Spoken Language Processing (ISCSLP)},
  pages={199--203},
  year={2024},
  organization={IEEE}
}

@inproceedings{saeki2022utmos,
  title={UTMOS: UTokyo-SaruLab System for VoiceMOS Challenge 2022},
  author={Saeki, Takaaki and Xin, Detai and Nakata, Wataru and Koriyama, Tomoki and Takamichi, Shinnosuke and Saruwatari, Hiroshi},
  booktitle={Proc. Interspeech 2022},
  pages={4521--4525},
  year={2022}
}

@inproceedings{serra2021sesqa,
  title={SESQA: semi-supervised learning for speech quality assessment},
  author={Serr{\`a}, Joan and Pons, Jordi and Pascual, Santiago},
  booktitle=icassp,
  pages={381--385},
  year={2021},
  organization={IEEE}
}

@inproceedings{kilgour2019frechet,
  title={Fr{\'e}chet Audio Distance: A Reference-Free Metric for Evaluating Music Enhancement Algorithms},
  author={Kilgour, Kevin and Zuluaga, Mauricio and Roblek, Dominik and Sharifi, Matthew},
  booktitle={Proc. Interspeech 2019},
  pages={2350--2354},
  year={2019}
}

@inproceedings{leng2021mbnet,
  title={MBNet: MOS prediction for synthesized speech with mean-bias network},
  author={Leng, Yichong and Tan, Xu and Zhao, Sheng and Soong, Frank and Li, Xiang-Yang and Qin, Tao},
  booktitle=icassp,
  pages={391--395},
  year={2021},
  organization={IEEE}
}

@inproceedings{vinay2022evaluating,
  title={Evaluating Generative Audio Systems and Their Metrics},
  author={Vinay, Ashvala and Lerch, Alexander},
  booktitle={Ismir 2022 Hybrid Conference},
  year={2022}
}

@inproceedings{majumder2024tango,
  title={Tango 2: Aligning diffusion-based text-to-audio generations through direct preference optimization},
  author={Majumder, Navonil and Hung, Chia-Yu and Ghosal, Deepanway and Hsu, Wei-Ning and Mihalcea, Rada and Poria, Soujanya},
  booktitle={Proceedings of the 32nd ACM International Conference on Multimedia},
  pages={564--572},
  year={2024}
}

@inproceedings{zhu2023ernie,
  title={ERNIE-Music: Text-to-Waveform Music Generation with Diffusion Models},
  author={Zhu, Pengfei and Pang, Chao and Chai, Yekun and Li, Lei and Wang, Shuohuan and Sun, Yu and Tian, Hao and Wu, Hua},
  booktitle={Proceedings of the 13th International Joint Conference on Natural Language Processing and the 3rd Conference of the Asia-Pacific Chapter of the Association for Computational Linguistics: System Demonstrations},
  pages={86--95},
  year={2023}
}

@article{liu2024audioldm,
  title={AudioLDM 2: Learning Holistic Audio Generation With Self-Supervised Pretraining},
  author={Liu, Haohe and Yuan, Yi and Liu, Xubo and Mei, Xinhao and Kong, Qiuqiang and Tian, Qiao and Wang, Yuping and Wang, Wenwu and Wang, Yuxuan and Plumbley, Mark D},
  journal=atraslp,
  volume={32},
  pages={2871--2883},
  year={2024},
  publisher={IEEE}
}

@inproceedings{ziv2024masked,
  title={Masked Audio Generation using a Single Non-Autoregressive Transformer},
  author={Ziv, Alon and Gat, Itai and Le Lan, Gael and Remez, Tal and Kreuk, Felix and Copet, Jade and D{\'e}fossez, Alexandre and Synnaeve, Gabriel and Adi, Yossi},
year={2024},
  booktitle={The Twelfth International Conference on Learning Representations}
}

@inproceedings{mittag2021nisqa,
  title={NISQA: A Deep CNN-Self-Attention Model for Multidimensional Speech Quality Prediction with Crowdsourced Datasets},
  author={Mittag, Gabriel and Naderi, Babak and Chehadi, Assmaa and M{\"o}ller, Sebastian},
  booktitle={Proc. Interspeech 2021},
  pages={2127--2131},
  year={2021}
}

@inproceedings{wang2024uncertainty,
  title={Uncertainty-Aware Mean Opinion Score Prediction},
  author={Wang, Hui and Zhao, Shiwan and Zhou, Jiaming and Zheng, Xiguang and Sun, Haoqin and Wang, Xuechen and Qin, Yong},
  booktitle={Proc. Interspeech 2024},
  pages={1215--1219},
  year={2024}
}

@article{wisnu2025improving,
  title={Improving Perceptual Audio Aesthetic Assessment via Triplet Loss and Self-Supervised Embeddings},
  author={Wisnu, Dyah AMG and Zezario, Ryandhimas E and Rini, Stefano and Wang, Hsin-Min and Tsao, Yu},
  journal={arXiv preprint arXiv:2509.03292},
  year={2025}
}

\end{document}